\documentstyle[12pt]{article}
\makeatletter
\newcommand{\beq}{\begin{equation}}
\newcommand{\eeq}{\end{equation}}
\newcommand{\beqy}{\begin{eqnarray}}
\newcommand{\eeqy}{\end{eqnarray}}

\def\zbar{\overline{z}}
\def\Zbar{\overline{Z}}
\def\thetabar{\overline{\theta}}
\def\Thetabar{\overline{\Theta}}
\def\mubar{\overline{\mu}}
\def\varphibar{\overline{\varphi}}
\def\partialbar{\overline{\partial}}
\def\alphabar{\overline{\alpha}}
\def\chibar{\overline{\chi}}

\def\STr{{\rm STr}}
\begin{document}
\begin{flushleft}
{\it  Yukawa Institute Kyoto}
\end{flushleft}
\begin{flushright}
YITP-96-62\\hep-th/9612031\\December 1996
\end{flushright}
\renewcommand{\thefootnote}{\fnsymbol{footnote}}
\vspace{0.5in}
\begin{center}\Large{\bf $Osp(1|2)$ Chern-Simons gauge theory\\
as 2D $N=1$ Induced Supergravity}\\
\vspace{1cm}
\normalsize\ Kiyoshi Ezawa\footnote[1]
{Supported by JSPS. E-mail address: ezawa@yukawa.kyoto-u.ac.jp .}
$\quad$and$\quad$ Atushi Ishikawa\footnote[2]{
Supported by JSPS. E-mail address: ishikawa@yukawa.kyoto-u.ac.jp .}
\vspace{0.5in}

        Yukawa Institute for Theoretical Physics \\
        Kyoto University, Kyoto 606-01, Japan\\
\vspace{0.1in}
\end{center}
\renewcommand{\thefootnote}{\arabic{footnote}}
\setcounter{footnote}{0}
\vspace{0.9in}
\baselineskip 16pt
\begin{abstract}
We demonstrate the close relationship between Chern-Simons gauge
theory with gauge group $Osp(1|2)$ and $N=1$ induced supergravity
in two dimensions. More precisely, the inner product of
the physical states in the former yields the partition function
of the latter evaluated in the Wess-Zumino supergauge.
It is also shown that the moduli space of
flat $Osp(1|2)$ connections naturally includes super
Teichm\"uller space of super Riemann surfaces.
\end{abstract}


\newpage

\baselineskip 17pt


It is important to quantize 2D induced (super)gravity
because it describes the dynamics of string world sheet
induced by the motion of (super)strings in the background
without critical dimensions\cite{polyakov}.
A number of interesting phenomena such as the fractal structure
have been clarified by vigorous works made in the light-cone gauge
\cite{KPZ}\cite{PZ} and in the conformal gauge\cite{DDK}\cite{DHK}.
A local expression of the generally covariant action for
2D induced bosonic gravity was obtained by Verlinde\cite{verlinde} 
in terms of the Beltrami coefficients.
More precisely, he showed that the partition function of the
induced 2D gravity is obtained from the inner product
of physical states in $SL(2,{\bf R})$ Chern-Simons gauge theory.
He also showed explicitly how we can extract the Teichm\"uller
parameters of Riemann surfaces from the holonomy of
flat $SL(2,{\bf R})$ connections.

In this paper we extend Verlinde's results to 2D
$N=1$ induced supergravity.
Namely, starting from $Osp(1|2)$ Chern-Simons gauge theory,
we obtain the super-covariant action for 2D $N=1$ induced
supergravity in the Wess-Zumino supergauge.
We also demonstrate that the moduli space of flat $Osp(1|2)$
connections yields super Teichm\"uller space of DeWitt
super Riemann surfaces with arbitrary spin structures.

Let us begin by briefly reviewing the Beltrami parametrization
of supervielbeins in 2D $N=1$ supergravity.
A DeWitt super Riemann surface
$S\Sigma$\cite{dewitt}
is a fiber bundle over a Riemann surface $\Sigma$
(with the coordinate reference
frame $(z,\zbar)$) whose fiber is a vector space
parametrized by two Grassmann odd coordinates $(\theta,\thetabar)$.
The supervielbein for the rigid superspace is given
by\footnote{We will use the convention of ref.\cite{WB} for
differential forms. As for the complex conjugation
of the product of two Grassmann odd variables,
we adopt the rule
$\overline{(\chi\eta)}=\overline{\chi}\: \overline{\eta}$.}
$e^{z}=dz+\theta d\theta$, $e^{\theta}=d\theta$
and their complex conjugates.
Their dual vectors are
$\partial=\frac{\partial}{\partial z}$,
$D=\frac{\partial}{\partial \theta}+\theta\frac{\partial}{\partial z}$
and their complex conjugates.
According to refs.\cite{10}\cite{grimm}, any supervielbein
$\{E^{A};\: A=+,-,\hat{1},\hat{2}\}$ which
represents Howe's superspace geometry\cite{howe}
can be written as follows:
\beqy
E^{+}&=&\rho e^{Z} \qquad,\qquad E^{-}=\overline{E^{+}}, \nonumber \\
E^{\hat{1}}&=&\sqrt{\rho}\{e^{\Theta}+\frac{1}{2}e^{Z}D_{\Theta}
\ln(\rho\overline{\rho})\}\quad,\quad E^{\hat{2}}=
\overline{E^{\hat{1}}}, \label{eq:vielbein}
\eeqy
where $\rho$ is the superconformal factor.
$e^{Z}\equiv dZ+\Theta d\Theta$,  $e^{\Theta}\equiv d\Theta$ and
their complex conjugates constitute a local basis of one-forms
on $S\Sigma$. Thier dual vectors are given by
$\partial_{Z}=\frac{\partial}{\partial Z}$,
$D_{\Theta}=\frac{\partial}{\partial\Theta}+\Theta
\frac{\partial}{\partial Z}$ and thier complex conjugates.

Because $(Z,\overline{Z})$ and $(\Theta,\Thetabar)$
are smooth scalar functions of
corrdinates $(z,\overline{z},\theta,\overline{\theta})$
which are Grassmann even and odd respectively,
we can expand $e^{Z}$ and $e^{\Theta}$ by the rigid basis
\beqy
e^{Z}&=&(e^{z}+e^{\zbar}H_{\zbar}^{z}+e^{\theta}H_{\theta}^{z}
+e^{\thetabar}H_{\thetabar}^{z})\Lambda\quad\mbox{ and C.C.},
\label{eq:beltrami} \\
e^{\Theta}&=&(e^{z}+e^{\zbar}H_{\zbar}^{z}+e^{\theta}H_{\theta}^{z}
+e^{\thetabar}H_{\thetabar}^{z})\tau
+(e^{\theta}H_{\theta}^{\theta}+e^{\zbar}H_{\zbar}^{\theta}
+e^{\thetabar}H_{\thetabar}^{\theta})\sqrt{\Lambda}\:
\mbox{ and C.C. }.\nonumber
\eeqy
Here $H_{M}^{N}$ $(M,N=z,\zbar,\theta,\thetabar)$ are called
Beltrami coefficients and $\Lambda$ and $\tau$ are called
integrating factors.
Owing to the structure equations $de^{Z}=-e^{\Theta}e^{\Theta}$
and $de^{\Theta}=0$, not all of these coefficients are independent.
It turns out that the Beltrami coefficients are local functionals
of $H_{\theta}^{z}$ and $H_{\thetabar}^{z}$
(and their complex conjugates)
and that $\tau$ is locally expressed by
$\Lambda$ and $H_{\theta}^{z}$. Moreover, the integrating factor
$\Lambda$ is subject to the equation:
\beq
\left[\overline{D}-\left(\frac{H_{\thetabar}^{\theta}}{
H_{\theta}^{\theta}}\right)D-\left(H_{\thetabar}^{z}-
\frac{H_{\thetabar}^{\theta}}{H_{\theta}^{\theta}}H_{\theta}^{z}
\right)\partial\right]\ln\Lambda=\partial H_{\thetabar}^{z}
-\frac{H_{\thetabar}^{\theta}}{H_{\theta}^{\theta}}
\partial H_{\theta}^{z}. \label{eq:IFEQ}
\eeq
The integrating factor $\Lambda$ is therefore regarded as
a nonlocal functional of $(H_{\theta}^{z},H_{\thetabar}^{z})$
which is uniquely determined up to superconformal redefinitions
of $(Z,\Theta)$. In consequence, 2D $N=1$ induced supergravity
is described by 20 local degrees of freedom which are
component fields of $(H_{\theta}^{z},H_{\thetabar}^{z},
H_{\thetabar}^{\zbar},H_{\theta}^{\zbar})$ and of 
$\rho\overline{\rho}$.
Instead of $\rho\overline{\rho}$ we will frequently use the
super-Liouville field
$
\Phi\equiv\ln(\rho\overline{\rho}\Lambda\overline{\Lambda})
=\phi+\theta\chi+\thetabar\overline{\chi}
+\theta\thetabar iF.
$
2D $N=1$ supergravity is invariant under graded local Lorentz
transformations:
$\rho\rightarrow e^{i\sigma}\rho,${}
$\overline{\rho}\rightarrow e^{-i\sigma}
\overline{\rho}$ ($\sigma$ is a real superfield).
The associated covariant derivative for a superfield $\Xi$ of
Lorentz weight\footnote{
For example, the Lortentz weights of $E^{\pm}$, $E^{\hat{1}}$ and
$E^{\hat{2}}$ are $\pm 1$, $+\frac{1}{2}$ and
$-\frac{1}{2}$, respectively.} $w$ is
given by${\cal D}\Xi=d\Xi+iw\Xi\Omega,$
where $\Omega$ is the graded spin connection
\beq
\Omega=-ie^{Z}\partial_{Z}\ln\overline{\rho}+ie^{\Zbar}
\partial_{\Zbar}\ln\rho-ie^{\Theta}D_{\Theta}\ln\overline{\rho}
+ie^{\Thetabar}D_{\Thetabar}\ln\rho .
\eeq
From this graded spin connection we can calculate the superfield
whose $\theta\thetabar$ component yields the
scalar curvature\cite{grimm}.
We will henceforth call this superfield the \lq\lq supercurvature''
$R_{3}$:
\beq
R_{3}=-2(\rho\overline{\rho})^{-1/2}D_{\Theta}D_{\Thetabar}
\ln(\rho\overline{\rho}). \label{eq:supercurvature}
\eeq

Because 2D $N=1$ supergravity should have symmetry under
reparametrizations of the super-coordinates $(z,\zbar,\theta,
\thetabar)$, there are a large number of choices to fix a gauge.
From now on we will consider a particular gauge-fixing
called the  Wess-Zumino (WZ) supergauge
\beq
H_{\theta}^{z}=0,\quad H_{\thetabar}^{z}=\thetabar\mu+\theta\thetabar
(-i\alpha). \label{eq:WZ}
\eeq
This gauge fixing condition is equivalent to the condition
\beqy
Z&=&Z_{0}(z,\zbar)+\theta\Theta_{0}\sqrt{\partial Z_{0}
+\Theta_{0}\partial\Theta_{0}}\quad\mbox{and C.C.},\nonumber \\
\Theta&=&\Theta_{0}(z,\zbar)+\theta\sqrt{\partial Z_{0}
+\Theta_{0}\partial\Theta_{0}}\quad\mbox{and C.C. }.\label{eq:WZ2}
\eeqy
A merit of the WZ supergauge is that all the component fields
(except the auxiliary field $F$) appear in the lowest
component of the supervielbein. Thus it is convenient
to introduce the zweibein $e^{\pm}(z,\zbar)$ and the
gravitino $\psi^{\alpha}(z,\zbar)$ $(\alpha=\hat{1},\hat{2})$:
\beqy
e^{+}&\equiv&E^{+}_{m}|dx^{m}=e^{\varphi}(dz+\mu d\zbar),
\quad e^{-}=\overline{e^{+}}, \nonumber \\
\psi^{\hat{1}}&\equiv&2E^{\hat{1}}_{m}|dx^{m}=
e^{\varphi/2}\{\chi(dz+\mu d\zbar)+i\alpha d\zbar\},
\quad \psi^{\hat{2}}=\overline{\psi^{\hat{1}}}.
\eeqy
Here we have used the notation $(x^{m})=(z,\zbar)$ and
the vertical bar denotes the $\theta=\thetabar=0$ component.
We have also used $\varphi$ to mean
the $\theta=\thetabar=0$ component of $\ln(\rho\Lambda)$.
We should note the relation $\varphi+\overline{\varphi}=\phi$.
The spin connection associated with the local Lorentz symmetry
is given by
\beq
\omega\equiv\Omega|=-i(dz+\mu d\zbar)\frac{(\partial-
\overline{\mu}\overline{\partial})\overline{\varphi}
-\overline{\partial}\overline{\mu}+\frac{i}{2}\overline{\alpha}
\overline{\chi}}{1-\mu\overline{\mu}}\quad +\mbox{ C.C. }. 
\eeq
As a consequence of the torsion
constraints\cite{howe}\cite{grimm} this spin connection satisfies
\beq
{\cal D}e^{+}\equiv de^{+}+ie^{+}\omega=
-\frac{1}{4}\psi^{\hat{1}}\psi^{\hat{1}}\quad \mbox{ and C.C. }.
\label{eq:torsion}
\eeq

$N=1$ supergravity in the WZ supergauge can be regarded as a minimal
superextention of the bosonic gravity in the sense that
the theory has residual gauge symmetry under
general coordinate transformations and  local SUSY
transformations generated by the spinor field parameter
$(\epsilon,\overline{\epsilon})$
\beq
\delta_{\mbox{{\tiny SUSY}}}
e^{+}=\frac{1}{2}\epsilon\psi^{\hat{1}},\quad
\delta_{\mbox{{\tiny SUSY}}}
\psi^{\hat{1}}={\cal D}\epsilon+\frac{1}{2}
\overline{\epsilon}(iFe^{-\phi/2})e^{+}\quad\mbox{ and C.C. }.
\label{eq:localSUSY}
\eeq

We are now ready to demonstrate the relationship
between $Osp(1|2)$ Chern-Simons gauge theory and
2D $N=1$ induced supergravity in the WZ supergauge.
For this purpose we first introduce the $Osp(1|2)$ connection
\beq
A\equiv-\omega J_{3}-i\lambda e^{a}J_{a}+\psi^{\alpha}Q_{\alpha},
\eeq
where $a=1,2$, and $(J_{i},Q_{\alpha})$
($i=1,2,3$ and $\alpha=\hat{1},\hat{2}$) are
generators of the $Osp(1|2)$ algebra:
\beq
[J_{i},J_{j}]=\epsilon_{ijk}J_{k},\quad
[J_{i},Q_{\alpha}]=-(\frac{\sigma_{i}}{2i})_{
\alpha}^{\beta}Q_{\beta}, \quad
\{Q_{\alpha},Q_{\beta}\}=\frac{i}{2}\lambda
(\frac{\sigma_{i}}{2i})_{\alpha}^{\gamma}\epsilon_{\gamma\beta}.
\eeq
It is possible to represent this connection $A$ by a $3\times 3$
matrix-valued one-form. If we define $e^{\pm}=-e^{2}\pm ie^{1}$
(and $J_{\pm}=-J_{2}\mp iJ_{1}$), the connection $A$ is written as
\beq
A=\left(\begin{array}{ccc}
\frac{i}{2}\omega & -\frac{i\lambda}{2}e^{-} & 
-\sqrt{\frac{i\lambda}{8}}\psi^{\hat{2}}\\
\frac{i\lambda}{2}e^{+} & -\frac{i}{2}\omega &
\sqrt{\frac{i\lambda}{8}}\psi^{\hat{1}} \\
\sqrt{\frac{i\lambda}{8}}\psi^{\hat{1}} &
\sqrt{\frac{i\lambda}{8}}\psi^{\hat{2}} & 0
\end{array}\right).\label{eq:osp(1,2)conn}
\eeq
We should remark that the bosonic part of this representation
yields the $SL(2,{\bf R})$ (or $SU(2)$) connection if the
parameter $\lambda$ is real (or pure imaginary),\footnote{
Note that this is true in the algebraic sense but not
literally. Namely, we have to perform the uitary transformation
$A\rightarrow A^{\prime}=e^{-\frac{\pi}{2}J_{1}}A
e^{\frac{\pi}{2}J_{1}}$
in order to make the bosonic part $SL(2,{\bf R})$-valued
in its original sense.
The whole connection $A^{\prime}$ then becomes
$Osp(1|2;{\bf R})$-valued.} corresponding
to the case of genus $g\geq 2$ (or $g=0$).

Next we calculate the curvature of the connection. We find
\beqy
{\cal F}&\equiv&dA+AA\:(\equiv{\cal F}^{i}J_{i}+
{\cal F}^{\alpha}Q_{\alpha})\nonumber \\
&=&-(d\omega+\frac{i\lambda^{2}}{2}e^{+}e^{-}-\frac{\lambda}{4}
\psi^{\hat{1}}\psi^{\hat{2}})J_{3} \nonumber \\
& &-\frac{i\lambda}{2}({\cal D}e^{+}+\frac{1}{4}
\psi^{\hat{1}}\psi^{\hat{1}})J_{+}
-\frac{i\lambda}{2}({\cal D}e^{-}+\frac{1}{4}
\psi^{\hat{2}}\psi^{\hat{2}})J_{-}\nonumber \\
& &+({\cal D}\psi^{\hat{1}}-\frac{i\lambda}{2}e^{+}
\psi^{\hat{2}})Q_{\hat{1}}+({\cal D}\psi^{\hat{2}}+
\frac{i\lambda}{2}e^{-}\psi^{\hat{1}})Q_{\hat{2}}.
\label{eq:osp(1,2)curvature}
\eeqy
From this expression we see immediately that ${\cal F}^{\pm}=0$
is equivalent to the torsion condition eq.(\ref{eq:torsion}).
Moerover, after somewhat tedious calculation,
it turns out that the equations ${\cal F}^{3}={\cal F}^{\hat{1}}
={\cal F}^{\hat{2}}=0$ together with the equation $F=\lambda
e^{\phi/2}$ yield the condition that the supercurvature in the WZ
supergauge is constant: \ $R_{3}=2i\lambda$.

Let us now consider the local gauge transformation:
\beq
\delta_{\zeta}A=-d\zeta+[A,\zeta],\quad
\zeta=\zeta^{3} J_{3}+\zeta^{a}J_{a}+\epsilon^{\alpha}Q_{\alpha}.
\eeq
The result of computation tells us that $\zeta^{3}$
generates local Lorentz transformations and that
the transformations generated by $\epsilon^{\alpha}$
is nothing but local SUSY transformations (\ref{eq:localSUSY})
if we set $F=\lambda e^{\phi/2}$. As for the transformations
generated by $\zeta^{a}$, it is closely related to diffeomorphisms.
This is because diffeomorphisms of the flat connection are
generated by gauge parameters of the form
$\zeta=\xi^{m}A_{m}$ and because we are only interested
in the case where the zweibein is nondegenerate. 

From the above considerations we expect that $Osp(1|2)$
Chern-Simons gauge theory (CSGT) on ${\bf R}\times\Sigma$
describes dynamics of super Teichmuller
space of super Riemann surfaces $S\Sigma$ in the WZ supergauge,
because any supervielbein is related to a
supervielbein with a constant supercurvature by a unique super-Weyl
transformation\cite{DPh}. We will see in the following
that the dynamics is in fact equivalent to that of
2D $N=1$ induced supergravity. We will use the
canonical quantization.

The action of $Osp(1|2)$ CSGT is given by
\beqy
S=\frac{k}{4\pi}\int_{{\bf R}\times\Sigma}\STr(\tilde{A}\tilde{d}
\tilde{A}+\frac{2}{3}\tilde{A}\tilde{A}\tilde{A}),
\eeqy
where $\tilde{A}=dtA_{t}+A$ and $\tilde{d}=dt\frac{\partial}
{\partial t}+d$ stand for the connection and the
exterior derivative defined on ${\bf R}\times\Sigma$, respectivrly.
$\STr$ denotes the invariant bilinear form in $Osp(1|2)$:
$$
\STr(J_{i}J_{j})=\delta_{ij},\quad \STr(Q_{\alpha}Q_{\beta})
=-\frac{i\lambda}{2}\epsilon_{\alpha\beta},\quad
\STr(J_{i}Q_{\alpha})=0.
$$
After a (3+1)-decomposition the action becomes
\beq
S=\frac{k}{4\pi}\int dt\int_{\Sigma}\STr[A\dot{A}+2A_{t}{\cal F}]
\equiv S_{kin}+S_{con},
\eeq
where $\dot{A}\equiv\frac{\partial}{\partial t}A$.
The second term $S_{con}$ yields the Gauss law constraint
${\cal F}=0$ and from the first term $S_{kin}$ we can read off
the symplectic structure.
In terms of the component fields $S_{kin}$ is written as
\beq
S_{kin}=\frac{k}{4\pi}\int dt\int_{\Sigma}
(-2\dot{\omega_{z}}dz\omega_{\zbar}d\zbar+\lambda^{2}\dot{e^{+}}
e^{-}+i\lambda\dot{\psi^{\hat{1}}}\psi^{\hat{2}}).
\eeq
Quantum commutation relations of the
canonical operators are obtained from
the operator version of $i$ times the Poisson brackets.
We will choose the polarization in which $\omega$ ($\equiv
\omega_{z}$), $e^{+}$ and $\psi^{\hat{1}}$
are diagonal. In this polarization the wavefunctionals are
holomorphic with respect to the complex structure equipped with
the space of super Riemann surfaces.
For our purpose  it is more convenient to
use the parameters $(\omega,\varphi,\mu,\chi,\alpha)$ as 
configuration variables.
Their canonical conjugate momenta are
\beqy
\pi_{\omega}&=&-\frac{k}{2\pi}\omega_{\zbar},\nonumber \\
\pi_{\varphi}&=&\frac{k}{4\pi}[\lambda^{2}e^{\phi}(1-\mu\mubar)+
\frac{i\lambda}{2}e^{\phi/2}\{\chi\chibar(1-\mu\mubar)+
i\mu\chi\alphabar+i\mubar\chibar\alpha-\alpha\alphabar\}],
\nonumber \\
\pi_{\mu}&=&-\frac{k}{4\pi}[\lambda^{2}e^{\phi}\mubar+i\lambda
e^{\phi/2}\chi(\mubar\chibar-i\alphabar)],\nonumber \\
\pi_{\chi}&=&\frac{ik\lambda}{4\pi}e^{\phi/2}[\chibar
(1-\mu\mubar)+i\mu\alphabar],\quad
\pi_{\alpha}=-\frac{k\lambda}{4\pi}(-\mubar\chibar+i\alphabar).
\eeqy
On quatization these conjugate momenta are represented by $-i$ times
the functional derivatives w.r.t. the associated configuration
variables: $\hat{\pi_{f}}=-i\frac{\delta}{\delta f}$
($f=\omega,\varphi,\mu,\chi,\alpha$).

The physical wavefunctional $\Psi[\omega,\varphi,\mu,\chi,\alpha]$
must satisfy the Gauss law constraints: 
$\hat{{\cal F}}\cdot\Psi=0$.
Instead of solving these constraints directly, we will solve
a proper set of their linear combinations
which is classically equivalent
to the set of Gauss law constraints when the zweibein is nondegenerate.
A convenient set is given by
\beqy
d^{2}z{\cal G}^{3}&\equiv&-{\cal F}^{3},\quad 
d^{2}z{\cal G}^{+}\equiv\frac{2i}{\lambda}e^{-\varphi}{\cal F}^{+},
\quad 
d^{2}z{\cal G}^{\hat{1}}\equiv e^{-\varphi/2}{\cal F}^{\hat{1}}-
\frac{\chi}{2}{\cal G}^{+},\nonumber \\
d^{2}z{\cal G}^{-}&\equiv&
-\frac{k\lambda}{2\pi}(e^{\varphi}{\cal F}^{-}
+\mubar e^{\varphibar}{\cal F}^{+})+\frac{k\lambda}{4\pi}e^{\phi/2}
(i\alphabar-\mubar\chibar){\cal G}^{\hat{1}}+\chi{\cal G}^{\hat{2}},
\nonumber \\
d^{2}z{\cal G}^{\hat{2}}&\equiv&-\frac{k\lambda}{4\pi}
[e^{\varphi/2}{\cal F}^{\hat{2}}-\frac{i\alphabar-\mubar\chibar}{2}
e^{\phi/2}{\cal G}^{+}].
\eeqy
We adopt the ordering in which the momenta are put
on the right of the coordinates. Then the operator version of
these constraints are written as
\beqy
\hat{{\cal G}^{3}}&=&\partialbar\omega-\frac{2\pi i}{k}\partial
(\frac{\delta}{\delta\omega})+\frac{2\pi}{k}
\frac{\delta}{\delta\varphi}, \nonumber \\
\hat{{\cal G}^{+}}&=&(\partialbar-\mu\partial)\varphi
-\partial\mu-i\omega\mu
+\frac{i}{2}\chi\alpha-\frac{2\pi}{k}\frac{\delta}{\delta\omega},
\nonumber \\
\hat{{\cal G}^{\hat{1}}}&=&-\frac{i\alpha}{2}(\partial\varphi+i\omega)
-i\partial\alpha+(\partialbar-\mu\partial-\frac{1}{2}\partial\mu)\chi
+\frac{2\pi i}{k}\frac{\delta}{\delta \chi}, \nonumber \\
\hat{{\cal G}^{-}}&=&(\partial\varphi+i\omega-\partial)
\frac{\delta}{\delta\varphi}-(\partialbar-\mu\partial-2\partial\mu)
\frac{\delta}{\delta\mu} \nonumber \\
& &\qquad+(\frac{3}{2}\partial\alpha+
\frac{1}{2}\alpha\partial)\frac{\delta}{\delta\alpha}
+\frac{1}{2}(\partial\chi-\chi\partial)\frac{\delta}{\delta\chi},\\
\hat{{\cal G}^{\hat{2}}}&=&i(\partialbar-\mu\partial-\frac{3}{2}
\partial\mu)\frac{\delta}{\delta\alpha}+\frac{i\alpha}{2}
\frac{\delta}{\delta\mu}
+(\frac{1}{2}\partial\varphi+\frac{i}{2}
\omega-\partial)\frac{\delta}{\delta\chi}+\frac{\chi}{2}
\frac{\delta}{\delta\varphi}.\nonumber
\eeqy

Now we can solve the constraint equations $\hat{{\cal G}^{I}}\cdot
\Psi=0$ ($I=3,+,-,\hat{1},\hat{2}$). The procedure is almost
parallel to that in the bosonic case\cite{verlinde}.
Namely, we first determine the
$\omega$-dependence of $\Psi$
by solving $\hat{{\cal G}^{+}}\cdot\Psi=0$
and then determine the $(\varphi,\chi)$-dependence by solving
$\hat{{\cal G}^{3}}\cdot\Psi=\hat{{\cal G}^{\hat{1}}}\cdot\Psi=0$.
The result is
\beqy
\Psi[\omega,\varphi,\mu,\chi,\alpha]&=&
\exp\{\frac{ik}{2\pi}(S_{O}[\omega,\varphi,\mu,\chi,\alpha]
+S_{L}[\varphi,\mu,\chi,\alpha])\}\Psi[\mu,\alpha],\nonumber \\
S_{O}[\omega,\varphi,\mu,\chi,\alpha]&=&
\int_{\Sigma}d^{2}z[-\frac{\mu}{2}\omega^{2}-i\omega\{
(\partialbar-\mu\partial)\varphi-\partial\mu+\frac{i}{2}\chi\alpha\}],
\nonumber \\
S_{L}[\varphi,\mu,\chi,\alpha]&=&\int_{\Sigma}d^{2}z
[-\frac{1}{2}\partial\varphi\partialbar\varphi+\mu(\frac{1}{2}
(\partial\varphi)^{2}-\partial^{2}\varphi) \nonumber \\
& &+\frac{1}{2}\chi
(\partialbar-\mu\partial)\chi
-\frac{i}{2}\chi\alpha\partial\varphi-i\chi\partial\alpha].
\eeqy
By substituting this form of the
wavefunctional, we can reduce the remaining constraints
$\hat{{\cal G}^{-}}\cdot\Psi=\hat{{\cal G}^{\hat{2}}}\cdot\Psi=0$
to the following equations for the functional
$\Psi[\mu,\alpha]$:
\beqy
\hat{{\cal V}}\cdot\Psi[\mu,\alpha]&=&0,\;
\hat{{\cal V}}\equiv-(\partialbar-\mu\partial-2\partial\mu)
\frac{\delta}{\delta\mu}+(\frac{3}{2}\partial\alpha+\frac{1}{2}
\alpha\partial)\frac{\delta}{\delta\alpha}+\frac{ik}{2\pi}
\partial^{3}\mu,\nonumber \\
\hat{{\cal S}}\cdot\Psi[\mu,\alpha]&=&0,\;
\hat{{\cal S}}\equiv(\partialbar-\mu\partial-\frac{3}{2}\partial\mu)
\frac{\delta}{\delta\alpha}+\frac{1}{2}\alpha\frac{\delta}{\delta\mu}
+\frac{ik}{2\pi}\partial^{2}\alpha.\label{eq:VW}
\eeqy
These equations are nothing but the Virasoro Ward identities
for a superconformal field theory.
These identities specify the transformation laws of superconformal
blocks under diffeomorphisms and local SUSY transformations\cite{3}.
A solution to these identities is known to be of the following form
\cite{3}
\beqy
\Psi[\mu,\alpha]&=&\exp(\frac{ik}{2\pi}S_{V}[\mu,\alpha])
\widetilde{\Psi}[\mu,\alpha],\nonumber \\
S_{V}[\mu,\alpha]&=&-\frac{1}{2}\int_{\Sigma}d^{2}z
\int d^{2}\theta H_{\thetabar}^{z}\partial D\ln\Lambda\nonumber \\
&=&-\frac{1}{2}\int_{\Sigma}d^{2}z\int d^{2}\theta
\frac{\overline{D}Z-\Theta\overline{D}\Theta}{\partial
Z+\Theta\partial\Theta}D\left(
\frac{\partial^{2}Z+\Theta\partial^{2}\Theta}{\partial
Z+\Theta\partial\Theta}\right),\label{eq:solution2}
\eeqy
where $(Z,\Theta)$ are given by eq.(\ref{eq:WZ2})
and $\widetilde{\Psi}[\mu,\alpha]$ is a functional of
Beltrami differentials which
is invariant under diffeomorphisms and local SUSY
transformations.
Thus we obtain the physical state of
$Osp(1|2)$ CSGT in our polarization:
\beq
\Psi[\omega,\varphi,\mu,\chi,\alpha]=
\exp\{\frac{ik}{2\pi}(S_{O}[\omega,\varphi,\mu,\chi,\alpha]
+S_{L}[\varphi,\mu,\chi,\alpha]+S_{V}[\mu,\alpha])\}
\widetilde{\Psi}[\mu,\alpha].\label{eq:solution}
\eeq

In order to discuss the dynamics of CSGT we have to consider
the inner product. Because the Hamiltonian
in CSGT is a linear combination of Gauss law
constraints, the transition amplitudes in CSGT
reduce to the inner products evaluated at a fixed time.
In our polarization, the physically relevant inner product which
yields the correct Hermitian conjugate condition is
\beqy
<\Psi_{1}|\Psi_{2}>&=&\int[d\omega de^{+}de^{-}
d\psi^{\hat{1}}d\psi^{\hat{2}}]\exp\left(-\frac{ik}{4\pi}
\int(-2\omega_{z} dz\omega_{\zbar}d\zbar+\lambda^{2}e^{+}e^{-}
+i\lambda\psi^{\hat{1}}\psi^{\hat{2}})\right) \nonumber \\
& &\qquad\times\overline{\Psi_{1}[\omega_{z},e^{+},\psi^{\hat{1}}]}
\Psi_{2}[\omega_{z},e^{+},\psi^{\hat{1}}].
\eeqy
If we substutute eq.(\ref{eq:solution}) for $\Psi_{1}$ and $\Psi_{2}$,
then the $\omega$ integration is easily performed and we are left with
\beq
<\Psi_{1}|\Psi_{2}>=
\int[de^{+}de^{-}d\psi^{\hat{1}}d\psi^{\hat{2}}]
\exp\left(\frac{ik}{2\pi}S_{GSC}^{(WZ)}\right)
\overline{
\widetilde{\Psi}_{1}[\mu,\alpha]}\widetilde{\Psi}_{2}[\mu,\alpha],
\label{eq:2Dsugra}
\eeq
where $S_{GSC}^{(WZ)}\equiv S_{L}+K[\mu,\mubar]+S_{V}[\mu,\alpha]-
\overline{S_{V}[\mu,\alpha]}+S_{\lambda}$. Here
$S_{L}$, $K[\mu,\mubar]$ and $S_{\lambda}$ stand for
the super-Liouville action, the super-extension of Verlinde's
local counterterm \cite{AGN} and the cosmological term
in the WZ supergauge, respectively:
\beqy
S_{L}&=&\int d^{2}z\left[\frac{-1}{2(1-\mu\mubar)}\{
(\partial\phi-\mubar\partialbar\phi+i\alphabar\chibar)
(\partialbar\phi-\mu\partial\phi-i\alpha\chi)-\frac{1}{2}
(\alpha\chi)(\alphabar\chibar)\}\right.\nonumber \\
& &+\frac{1}{2}\chi(\partialbar-\mu\partial)\chi+\frac{1}{2}
\chibar(\partial-\mubar\partialbar)\chibar-i\chi\partial\alpha
+i\chibar\partialbar\alphabar \nonumber \\
& &\left.+\frac{\partial\mu-\mu\partialbar\mubar}{1-\mu\mubar}
(\partial\phi+\frac{i}{2}\alphabar\chibar)
+\frac{\partialbar\mubar-\mubar\partial\mu}{1-\mu\mubar}
(\partialbar\phi-\frac{i}{2}\alpha\chi) \right], \nonumber \\
K[\mu,\mubar]&=&\int d^{2}z\frac{-1}{1-\mu\mubar}
\left(\partial\mu\partialbar\mubar-\frac{\mubar}{2}(\partial\mu)^{2}
-\frac{\mu}{2}(\partialbar\mubar)^{2}\right), \nonumber \\
S_{\lambda}&=&-\frac{\lambda^{2}}{2}\int e^{+}e^{-}
-\frac{i\lambda}{2}\int\psi^{\hat{1}}\psi^{\hat{2}}.
\eeqy
After a lengthy and tedious calculation
we see that the integrand in eq.(\ref{eq:2Dsugra}) is
indeed invariant under diffeomorphisms and local SUSY
transformations (\ref{eq:localSUSY}).
The exponent $S_{GSC}^{(WZ)}$ therefore
yields a natural super-extension of the
generally covariant action for 2D induced bosonic gravity.
Thus we can conclude that the inner product of the
physical states in $Osp(1|2)$ CSGT, eq.(\ref{eq:2Dsugra}),
gives rise to the partition function (or transition amplitudes)
of 2D $N=1$ supergravity in the WZ supergauge.

Next we consider the observables.
In CSGT on a manifold
with topology ${\bf R}\times\Sigma$, the connection $A$ on $\Sigma$
is always constrained to be flat, and only global objects
such as Wilson loops or holonomies are relevant.
In the following we will demonstrate that the space of
holonomies constructed from the flat $Osp(1|2)$ connections $A$ is
closely related to super Teichm\"uller space
of DeWitt super Riemann surfaces.
We restrict our discussion to the case of genus $g\geq2$
(i.e. $\lambda$: real).

According to the uniformization theorem for super Riemann surfaces
\cite{crane}, any genus $g\geq2$ super Riemann surface of Dewitt type
\cite{dewitt} (with a constant supercurvature) is represented as
a quotient of the super-upper-half plane $SH$
equipped with the super-Poincar\'e geometry
by a discrete hyperbolic subgroup of $Osp(1|2;{\bf R})$.
This implies that super Teichm\"uller space is represented as
the set of homomorphisms of $\pi_{1}(\Sigma)$
into $Osp(1|2;{\bf R})$ modulo overall conjugations.
What we have to examine is whether the holonomy of a flat
$Osp(1|2;{\bf R})$ connection directly yields such a homomorphism.
This is achieved by extending the result of the bosonic
case\cite{verlinde} to the case of supergravity.

We first note that the canonical supervielbein
with the constant supercurvature $R_{3}=2i\lambda$,
i.e. the super-Poincar\'e geometry,
is characterized by the conformal factor
\beq
\rho=\frac{2}{\lambda}\frac{e^{i\gamma(z,\zbar,\theta,\thetabar)}}{
Z-\Zbar-\Theta\Thetabar},
\eeq
where $\gamma(z,\zbar,\theta,\thetabar)$
is an arbitrary phase factor which is real. In the WZ supergauge
the corresponding zweibein, gravitino and spin connection are
\beqy
e^{+}&=&\frac{2}{\lambda}
\frac{e^{i\gamma_{0}(z,\zbar)}}{\Xi_{0}}e^{Z_{0}},\quad
\psi^{\hat{1}}=2\left(\frac{2}{\lambda}
\frac{e^{i\gamma_{0}(z,\zbar)}}{\Xi_{0}}\right)^{1/2}
\left[e^{\Theta_{0}}-e^{Z_{0}}\frac{\Theta_{0}-\Thetabar_{0}}{
\Xi_{0}}\right]\quad\mbox{ and C.C. }, \nonumber \\
\omega&=&\frac{i}{\Xi_{0}}
(dZ_{0}+d\Zbar_{0}+
\Thetabar_{0}d\Theta_{0}+\Theta_{0}d\Thetabar_{0})
-d\gamma_{0}(z,\zbar),
\eeqy
where we have used the notation $\Xi_{0}=Z_{0}-\Zbar_{0}-
\Theta_{0}\Thetabar_{0}$,
$e^{Z_{0}}=dZ_{0}+\Theta_{0}d\Theta_{0}$,
$e^{\Theta_{0}}=d\Theta_{0}$ and $\gamma_{0}\equiv\gamma|$.
Substituting these into
eq.(\ref{eq:osp(1,2)conn}), we find that
the $Osp(1|2;{\bf R})$ connection
$A$ is locally expressed as a pure gauge $A=-g^{-1}dg$ with
\beq
g=\left(\begin{array}{ccc}
e^{\frac{i\gamma_{0}}{2}+\frac{\pi i}{4}}Z_{0}\sqrt{\Xi_{0}}^{-1} &
e^{-\frac{i\gamma_{0}}{2}-\frac{\pi i}{4}}\Zbar_{0}\sqrt{\Xi_{0}}^{-1} &
(\Thetabar_{0}Z_{0}-\Theta_{0}\Zbar_{0})(\Xi_{0})^{-1} \\
e^{\frac{i\gamma_{0}}{2}+\frac{\pi i}{4}}\sqrt{\Xi_{0}}^{-1} &
e^{-\frac{i\gamma_{0}}{2}-\frac{\pi i}{4}}\sqrt{\Xi_{0}}^{-1} &
(\Thetabar_{0}-\Theta_{0})(\Xi_{0})^{-1} \\
-e^{\frac{i\gamma_{0}}{2}+\frac{\pi i}{4}}
\Theta_{0}\sqrt{\Xi_{0}}^{-1} &
-e^{-\frac{i\gamma_{0}}{2}-\frac{\pi i}{4}}
\Thetabar_{0}\sqrt{\Xi_{0}}^{-1} &
1-\Theta_{0}\Thetabar_{0}(\Xi_{0})^{-1}
\end{array}\right).
\eeq
Because the connection $A$ does not have to be
pure gauge globally, the $Osp(1|2)$-valued field
$g(z,\zbar)$ is not necessarily single-valued on $\Sigma$.
When one goes around any  non-contractible loop $\beta$ on the
surface $\Sigma$, $g(z,\zbar)$ in general transforms as 
\beq
g\longrightarrow h[\beta]\cdot g, \qquad
h[\beta]=\left(\begin{array}{ccc}
a & b & b\delta+a\epsilon \\
c & d & d\delta+c\epsilon \\
\delta & -\epsilon & 1+\delta\epsilon
\end{array}\right),
\eeq
where $a,b,c,d\in{\bf R}$ are Grassmann even constants,
$\overline{\delta}=\delta$ and $\overline{\epsilon}=\epsilon$
are Grassmann odd constants, and
the relation $ad-bc=1-\delta\epsilon$ holds.
This $h[\beta]$ is nothing but the holonomy of the $Osp(1|2;{\bf R})$
connection $A$ around the loop $\beta$.
This left  multiplication induces the following transformation
of  $(Z_{0},\Zbar_{0},\Theta_{0},\Thetabar_{0},\gamma_{0})$:
\beqy
Z_{0}&\longrightarrow&\frac{aZ_{0}+b}{cZ_{0}+d}+
\frac{\Theta_{0}(-\delta Z_{0}+\epsilon)}{(cZ_{0}+d)^{2}},\quad
\Theta_{0}\longrightarrow\frac{-\delta Z_{0}+\epsilon}{cZ_{0}+d}
+\frac{\Theta_{0}}{cZ_{0}+d}\quad\mbox{ and C.C. },\nonumber \\
\gamma_{0}&\longrightarrow&\gamma_{0}-i\ln\left(
\frac{cZ_{0}+d+\Theta_{0}(d\delta+c\epsilon)}{
c\Zbar_{0}+d+\Thetabar_{0}(d\delta+c\epsilon)}\right).
\eeqy
In the WZ supergauge, the transformation of $(Z,\Zbar,\Theta,
\Thetabar)$ can be read off from that of $(Z_{0},\Zbar_{0},
\Theta_{0},\Thetabar_{0})$. In the present case both transformations
coincide. Namely, we find
\beq
Z\longrightarrow\frac{aZ+b}{cZ+d}+
\frac{\Theta(-\delta Z+\epsilon)}{(cZ+d)^{2}},\quad
\Theta\longrightarrow\frac{-\delta Z+\epsilon}{cZ+d}
+\frac{\Theta}{cZ+d}\quad\mbox{ and C.C. }.
\eeq
This is the super-M\"obius transformation which plays
the essential role in the uniformization theorem.
If we regard $(Z,\Zbar,\Theta,\Thetabar)$ as a complex coordinate
system which maps the super-Riemann surface $S\Sigma$ into
the super-upper-half plane $SH$, the holonomy group of the flat
$Osp(1|2;{\bf R})$ connection $A$ turns out to be identical to
the discrete group by which $SH$ is divided out.
Thus we have explicitly shown the close relationship between the moduli
space of flat $Osp(1|2;{\bf R})$ connections
and super Teichm\"uller space.

Here we comment on spin structures.
A spin structure specifies whether or not the fermionic coordinate
$\Theta$ flips its sign when one goes around each cycle $\beta$
on $\Sigma$. This is determined by the signature of $d$ (or $c$
if $d=0$) in the $Osp(1|2;{\bf R})$ holonomy $h[\beta]$.
In the bosonic case, we cannot distinguish the difference
between the identity ($a=d=1$, $b=c=\delta=\epsilon=0$) and the inversion
($a=d=-1$ and $b=c=\delta=\epsilon=0$) and thus
a  M\"obius transformation is isomorphic to an element of $PSL(2,{\bf R})$.
In the super case, on the other hand,
the inversion is essentially different from the identity
because the former yields the spin stricture which
is distinct from that of the latter. Owing to this property,
a choice of $Osp(1|2;{\bf R})$
holonomy naturally specifies a unique spin structure.

To summarize the results we have seen that $Osp(1|2)$
Chern-Simons gauge theory describes the dynamics of
2D $N=1$ induced supergravity in the Wess-Zumino supergauge.
Physical inner products of $Osp(1|2)$ CSGT yields the
partition function (or transition amplitudes)
of 2D $N=1$ quantum supergravity.
From the holonomy of the $Osp(1|2)$ connection we can extract
the spin structure and the super Teichm\"uller parameters of
the super Riemann surface $S\Sigma$.

Is there any possibility of exploiting these consequences?
In this paper we have used the polarization in which
$(\omega_{z},e^{+},\psi^{\hat{1}})$ are diagonal
and obtained the partition function of 2D $N=1$ induced
supergravity. On the other hand, in the holomorphic
polarization in which $A_{z}$ is diagonal, it is well known that
the physical wavefunctional is given by the exponential
of the $Osp(1|2)$ WZNW action\cite{EMB}
in the case of a trivial topology.\footnote{
Exploiting this fact we can show that the light-cone gauge action
for $N=\frac{1}{2}$ supergravity ($S_{V}[\mu,\alpha]$ in
eq.(\ref{eq:solution2}))is related with the {\em Borel gauged}
$Osp(1|2)$ WZNW action through a Legendre transform. This establishes
the result obtained in ref.\cite{sabra} from a different viewpoint.}
Using the Polyakov-Wiegmann identity\cite{PW}, the inner
product in the holomorphic polarization reduces to
the partition function of the twisted
$Osp(1|2;{\bf C})/Osp(1|2;{\bf R})$ WZNW model.
Because the $G^{{\bf C}}/G$ WZNW model is a conformal field theory
which is well studied\cite{gawedzki}, this may help the description of
2D induced (super)gravity.
Before using such a description, however,
we will have to establish the direct relationship
between the $G^{{\bf C}}/G$ WZNW model (with $G=SL(2,{\bf R})$
or $Osp(1|2;{\bf R})$) and 2D induced (super)gravity.
This is not so easy and is left to the future investigation.

As a by-product, we have obtained the generally super covariant action
$S_{GSC}^{(WZ)}$ in the WZ supergauge. Actually this action can be
obtained from the expression
\beq
S_{GSC}=\int d^{2}Zd^{2}\Theta\left[-\frac{1}{2}D_{\Theta}
\ln(\rho\overline{\rho})D_{\Thetabar}\ln(\rho\overline{\rho})
+2i\lambda\sqrt{\rho\overline{\rho}}\right]
\eeq
which is manifestly invariant under reparametrizations of
super-coordinates:
$(z,\zbar,\theta,\thetabar)\rightarrow
(z^{\prime},\zbar^{\prime},\theta^{\prime},\thetabar^{\prime}).$
This $S_{GSC}$ is therefore regarded as a local expression
of the generally super covariant action for 2D $N=1$ induced
supergravity. Using this $S_{GSC}$ as a startpoint,
we can obtain the action in an arbitrary gauge simply by
fixing the gauge. This may open a way to make a further breakthrough
in 2D quantum supergravity.

\noindent {\large\bf Acknowledgments}

We would like to thank Prof. M. Ninomiya and Prof. Y. Matsuo
for warmful encouragements and careful readings of the manuscript.
We are also grateful to Dr. K. Sugiyama 
for useful discussions.

\end{document}